# Functional brain networks: great expectations, hard times, and the big leap forward

David Papo[1]*, Massimiliano Zanin[2,3], José Angel Pineda-Pardo[1], Stefano Boccaletti[4] and Javier M. Buldú[1,5]

[1]*Center for Biomedical Technology, Universidad Politécnica de Madrid, Madrid, Spain*
[2] *Faculdade de Cîencias e Tecnologia, Departamento de Engenharia, Electrotécnica, Universidade Nova de Lisboa, Lisboa, Portugal*
[3]*Innaxis Foundation & Research Institute, Madrid, Spain*
[4]*Istituto dei Sistemi Complessi, CNR, Florence, Italy*
[5]*Universidad Rey Juan Carlos, Móstoles, Spain*
*Correspondence: papodav@gmail.com

Many physical and biological systems can be studied using complex network theory, a new statistical physics understanding of graph theory. The recent application of complex network theory to the study of functional brain networks generated great enthusiasm as it allows addressing hitherto non-standard issues in the field, such as efficiency of brain functioning or vulnerability to damage. However, in spite of its high degree of generality, the theory was originally designed to describe systems profoundly different from the brain. We discuss some important caveats in the wholesale application of existing tools and concepts to a field they were not originally designed to describe. At the same time, we argue that complex network theory has not yet been taken full advantage of, as many of its important aspects are yet to make their appearance in the neuroscience literature. Finally, we propose that, rather than simply borrowing from an existing theory, functional neural networks can inspire a fundamental reformulation of complex network theory, to account for its exquisitely complex functioning mode.

## 1. Introduction

Characterizing how the brain organizes its activity to carry out complex tasks is highly non trivial. While early neuroimaging and electrophysiological studies typically aimed at identifying patches of task-specific activation or local time-varying patterns of activity, there has soon been consensus that task-related brain activity has a temporally multiscale, spatially extended character, as networks of coordinated brain areas are continuously formed and destroyed [1,2].

Up until recently, though, the emphasis of functional brain activity studies has been on the identity of the particular nodes forming these networks, and on the characterization of connectivity metrics between them [3], the underlying covert hypothesis being that each node, constituting a coarse-grained representation of a given brain region, provides a unique contribution to the whole. Thus, functional neuroimaging initially integrated the two basic ingredients of early neuropsychology: localization of cognitive function into specialized brain modules and the role of connection fibres in the integration of various modules.

Lately, brain structure and function have started being investigated using complex network theory, a statistical mechanics understanding of an old branch of pure mathematics: graph theory [4]. Graph theory allows endowing networks with a great number of quantitative properties [5,6], thus vastly enriching the set of objective descriptors of brain structure and function at neuroscientists' disposal.

However, in spite of a great potential, the results have so far not entirely met the expectations in that complex network theory has not yet given rise to a major breakthrough, has mainly been used to achieve descriptive goals, and has not yet had an impact on the way neurological or psychiatric pathologies are treated.

In this paper, we discuss possible reasons behind the current state of affairs and point to directions where graph theory could fruitfully be employed. In particular, we illustrate how complex network theory is used to describe functional brain activity, suggest alternatives to current practices, but also propose ways in which it could achieve further fundamental objectives, from classifying, to modelling, forecasting and even controlling brain activity.

## 2. Great expectations: graph theory's revolution

Complex network theory is not a mere additional set of tools in the neuroscientists' bag of tricks. Rather, it constitutes a major turning point, both conceptual and methodological.

### (a) A new paradigm for brain function

At a conceptual level, the complex network approach represents a paradigm shift from a computer-like to a complex system approach to the brain [7]. In the former approach, as is the case of computers, the brain is a collection of heterogeneous parts where function can be traced back to the computations carried out at well-defined locations and to the transport of their output from one location to another. At the system-level of investigation of standard non-invasive neuroimaging techniques, modelling typically involves a small number of units.

The huge number of neurons ($\sim 10^{11}$) and synapses ($\sim 10^{15}$) [8] suggests that the brain is better modelled as a complex system [9], capable of generating a vast repertoire of macroscopic patterns of collective behaviour with distinctive temporal, spatial or functional structures.

While modelling macroscopic behaviour in terms of only a few degrees of freedom as in the former approach likely represents a drastic reduction, the sheer number of interacting parts makes it unfeasible to study the brain's macroscopic functional properties by explicitly modelling each of its degrees of freedom. Statistical mechanics provides a framework for describing how these macroscopic patterns may result from the interactions of a multitude of microscopic individual entities [10]. Macro and microscopic scales are not absolute ones, but depend on the range of scales afforded by the experimental techniques used to observe brain activity, as well on the coarse-graining level used in data analysis. For instance, microscopic entities could be single neurons or neuronal micro-columns ($\sim 10^2$ neurons) or even neuronal populations comprising hundreds or thousands of micro-columns, etc.

The statistical mechanics approach underlying complex network theory [11] allows conceiving of macroscopic brain function as emerging in a non-trivial way from the interactions of a vast number of microscopic neural units. The networks formed by these interactions are endowed with properties which do not depend on those of their constituent nodes: neither particular nodes, nor particular links have, at least *prima facie*, an identifiable role in determining network properties. These are instead essentially statistical in nature. Ultimately, observable functional abilities are but the macroscopic output of the renormalization of neural fluctuations at microscopic scales.

### (b) A new way of describing brain activity

Both at rest and during the execution of cognitive tasks, the brain produces complex fluctuations at many spatial and temporal scales. Finding good collective variables describing such an inherently multiscale spatially-extended system's function is possibly the most impervious task facing neuroscientists.

Endowing brain activity with a network representation allows applying a set of mathematical tools, ultimately yielding valuable information on the collective behaviour of the brain.

#### (i) From important parts to general organizing principles

One of the main objectives of neuroscientists is typically identifying key brain regions responsible for certain observed behaviours. With complex network theory this can be accomplished at various scales. For instance, it is possible to identify and quantify the role played by the most basic actors of the network, i.e. nodes and links [12], and the extent to which they are playing a leading role. In turn, one can examine whether well-connected nodes display a specific connectivity, known as *rich club*, characterized by a tendency to denser connectivity than that of nodes with fewer connections [13].

The importance of a node in a network can be measured in various ways using *centrality* metrics [6]. Centrality may refer to a *leading node* of a brain region or to the *main connector* between different regions [12], and can be quantified in terms of local properties, e.g. the number (or weight) of connections, or global properties, e.g. the number of shortest paths connecting any pair of nodes crossing a given node. Correspondingly, various centrality measures, e.g. node degree, betweenness [14], or eigenvector centrality [15], have been proposed, each quantifying different ways in which some nodes of a network can be thought of as central.

Complex network theory allows going one step further and investigating general organizing principles at all scales, reflecting the fact relevant aspects of functional brain activity, such as information storing, may be either local, or non-locally spread across widely separated units.

At a global level, neither random nor regular lattices seem to constitute an adequate description of functional brain networks. Instead, it has been shown that these networks have small-world (SW) structure, indicating that any two vertices in the network can be connected through just a few links and, at the same time, that local connectivity is much denser than that of networks where connections are made at random [16]. Functional brain networks have also been reported to be *scale-free* [17,18], indicating a non negligible power-law probability $p(k)$ of finding nodes with a high number of connections $k$ [19].

In addition, functional connectivity has also been shown to be *assortative* [17], i.e. its nodes tend to form groups with nodes having a similar number of connections [20].

Within this global organization, a modular structure [21] has been highlighted. A *core-periphery* organization, where highly connected nodes form a stable dense core, surrounded by a periphery, composed of low-degree nodes with a time-varying connectivity can also occasionally be identified [22], providing insight into the way functional modules are connected with each other.

Complex network theory also allows characterizing non random patterns that are neither global nor local. Numerous mesoscale topological structures, possibly representing functionally relevant units, have been identified and their interactions assessed, using *ad hoc* detection algorithms [23,24]. These include for instance connectivity patterns between nodes that are overrepresented in the network known as *motifs* [25], or larger functionally coupled regions known as *community structures* [26].

Not only does complex network theory afford a description of brain activity at multiple scales, but it also helps unveiling various aspects of the relations between them. Various studies highlighted that the brain shows prominent hierarchical structure, with modules themselves containing other modules [26]. Zooming in and out of brain functional activity reveals a complex fractal structure, showing both self-similarity [27] and self-dissimilarity [28]. Interestingly, these global properties are associated with some mesoscale properties such as assortativity, with hubs in fractal and non fractal structures respectively repelling or attracting each other [29].

Complex network theory also allows quantifying the inherent trade-off between the emergence of segregated specialized modules, stemming from the need for fast and reliable responses to changes in the environment, and integrated global coherent activity, necessary for the binding of complex information and the formation of adaptive responses [30], and evaluating the extent to which this balance is optimized [31,32].

The fact that some of these properties have been found in systems that are very different from the brain, suggests that they may have a *universal* character [33], and may possibly be grouped into *universality classes*, identifying common interaction rules over and above the microscopic details of each particular system, which can be treated as irrelevant, as they disappear when getting rid of details and observing the system at increasingly longer length scales.

Appropriate null models facilitate quantitative network evaluation. A random network topology with the same degree distribution as the original network can be used as a null model [34].

### (ii) From structure to dynamics to function

The fact that the same methodology can be applied to networks of different nature is an advantage when delineating the relationship between anatomy, dynamics and function. For example, the topological properties of anatomical and functional networks have been compared with the aim of detecting influences or constraints of one network on the other [35-37]. Although difficult to compare, due to dissimilar origins and to a different number of nodes and links, these networks share some topological properties, such as the SW structure [38-40]. Several studies have compared functional networks at rest with structural connections, reporting high correspondence between both types of networks in regions with higher density of anatomical connections, while high variability in the functional correlation was associated with scarce anatomical connectivity [35-37].

Anatomical brain networks have been used as a benchmark to test dynamical models of spontaneous brain activity and how dynamics is affected by structure [35,37,41-49; see also 50,51 for recent reviews on the topic]. While the way resting connectivity relates to the anatomical connectivity remains an open question, the correlation structure of resting functional connectivity as measured by the slow spontaneous BOLD fluctuations was shown to relate to the underlying anatomical circuitry as obtained by diffusion tensor spectrum imaging [35,36]. This was taken to suggest that resting state networks arise from correlations of neuronal noise between brain areas that are coupled by the underlying anatomical connectivity [49].

However, at time scales faster than those of BOLD fluctuations, the relationship between functional and structural networks is far from clear [35,51]. In general, what sort of boundary condition anatomy plays, and at what (spatial and temporal) scales this constraint cannot be neglected remains largely unknown.

### (iii) Treating the brain as a biophysical object

In the complex system approach, the brain is thought of as a thermodynamical system, subject to energy costs and constraints, entropy barriers and information flows across its boundaries [8].

A wide range of nested, hitherto unaddressed theoretical and experimental questions arise naturally. For instance, how efficiently does the brain perform the functions it is supposed to carry out, under the constraints it is facing? How does it withstand external perturbations?

The way the human brain organizes its structure and function can be understood as the result of the constraint optimization process faced by any physical and biological system such as electronic devices or communications networks, and shows that similar principles of resource allocation can be found in many physical and biological systems [8]. These questions can be addressed by examining the topological and dynamical network properties. Far from being mere fancy mathematical descriptions of a system, these properties have important implications for the system's functioning [52].

The particular structure of connections has important consequences for the information processing capacities. The ability to process and propagate signals between nodes is for example affected by whether networks possess branching or loop-like features [53]. It also affects the efficiency and robustness of networks. For instance, several studies have suggested that the small-world organization of functional brain activity favours high communication efficiency for a low wiring cost [4]. It has also been shown that the presence of rich-club organization provides important information on network properties such as hierarchal structure, modularity and resilience [54,55]. The SW structure is commonly associated with an efficient organization of the brain, compatible with a simultaneous integration and segregation of information through the network [19]. On the other hand, scale-free networks are highly resistant against random failures [18,56], though extremely fragile to attacks targeting their most connected nodes [57]. Similarly, the presence of degree-degree correlations affects the tendency to separate into distinct groups, as well as network synchronizability and vulnerability to attacks. Specifically, assortative networks facilitate the spread of information over the network [20], are less vulnerable to attacks, but are more difficult to synchronize [58], and show a stronger resistance to disintegrate into different groups [59] than disassortative networks [20].

In addition, networks with heterogeneous components and modularity tend to have adaptive capacity, adjusting gradually to change. In highly connected networks, on the contrary, local losses tend to be withstood until the system reaches a critical stress level at which it collapses [61].

Network scientists have striven to translate some of these concepts into network measurable variables. For example, from an information transfer point of view, it is possible to quantify a network's ability to transmit a message in an efficient way, i.e. with the shorter number of steps between the sender and the receiver [52]. It is also possible to understand how network parameters are affected by targeted attacks or random failures, thus quantifying the vulnerability and robustness of functional networks [57].

Functional networks' potential for coordinating the dynamics of their nodes and proneness to synchronize could also be measured using *synchronizability* [62], a property that can be evaluated from the spectral properties of the network [63].

### (iv) Characterising functional brain disease and cognitive function

Ultimately, complex network theory would seem to allow characterizing how all these properties of functional networks of healthy brains are modulated under various experimental conditions, e.g. sensory stimulations, motor or cognitive tasks [19], and by neurological or psychiatric pathology, e.g. epilepsy [64,65], traumatic brain injury [66], brain tumours [67], mild cognitive impairment [42,68,69], Alzheimer's disease [70,71] or schizophrenia [72,73].

While it is not clear to what extent the execution of cognitive tasks influences global topological parameters or synchronizability [19], the fine structure of functional connectivity can be taken to reflect that of functional modules. For instance, the core-periphery organization was found to reflect the ability of subject to learn a motor skill, with participants showing a larger separation between core and periphery learning better than individuals with a smaller separation [22].

On the other hand, whether or not they stem from focal, spatially localized damage, neurological and psychiatric pathologies seem to affect the overall functional network structure, from global organization down to meso- and microscopic network scales [74]. Significant global changes may for instance involve loss of small-worldness, with networks becoming more random [42]. Pathology-related changes are also found both in the mesoscale structure of functional activity, e.g. in the community structure of the networks [26], and at the microscale, where the role played by single nodes may be altered [75]. For instance, network hubs appear to be especially vulnerable to brain disease [18,76], which appears to reshuffle the ranking of node centralities in the network [66,77].

## 3. Hard times

The introduction of fundamentally different concepts and tools to a new field is a path often plagued with pitfalls, typically coming under the guise of *over-*, *under-* and *mis-*application and interpretation.

In fact, the enthusiastic resort to any new method, particularly when imported from other fields, may on one hand lead into disregarding some of its assumptions and limitations and, on the other hand, lead to the (often improper) isolation of those parts of the method that are more readily amenable to the pillar concepts of pre-existing ones, while other parts are only resorted to at later stages, regardless of their possible scope in the field.

Neuroscience is not being spared these various problems, as stumbling blocks lurk at all levels, from the very domain of complex network theory's applicability, to the crucial choices made to build networks from empirical neuroimaging data, to the definition of network properties and their interpretation, and the principled discrimination of the most important ones.

### (a) Applicability of complex networks theory

The standard formulation of the statistical physics approach to graph theory implies a substantial equivalence of all its constituent nodes [78]. While the underlying substrate that each node is taken to represent may differ from one node to another, differences are generally supposed to be irrelevant.

However, from a statistical mechanics view-point, at the system-level network representation typical of non-invasive neuroimaging techniques, the brain can be thought of as a disordered system, with pronounced anatomical and physiological heterogeneity, and functional modularity.

In the presence of strong disorder and inhomogeneity, and complex structure-function relationships, the degree of coarse-graining of the system crucially determines the meaningfulness of a network representation. For a given spatial and temporal resolution, genuine property emergence predicated by the statistical mechanics approach may for instance not apply to the whole brain, but only to specific parts.

As a result, the node equivalence doesn't hold *prima facie*, leading to some fundamental questions: when does a brain network cease to be a complex network and start being a mere collection of nodes, with network properties reducing to simple connectivity? Is there a particular observation scale at which this occurs?

### (b) Building functional networks

Explaining functional brain activity in terms of objectively quantifiable functions of observed connectivity would seem to address one of the most fundamental concerns vexing neuroscientists, particularly those interested in brain functional activity.

However, complex network theory is neutral as to the way a network is reconstructed from empirical data. Identifying nodes, establishing links according to some relationship between them, deciding which links are significant, and, once network properties are computed, using them to characterize the network are all steps involving somehow arbitrary choices with often covert underlying assumptions, and far-reaching nested consequences. See [9;79-81] for recent critical reviews on the topic.

#### (i) Identifying nodes

Identifying nodes supposes that the studied system can meaningfully be decomposed into discrete structureless parts. This reduction is not trivial when dealing with systems of largely unknown organization and dynamics.

Depending on the technique used to record brain activity, the main issue may be the extent to which sensors sample the underlying dynamical system or how to best segment the space.

Studies using electrophysiological techniques such as electro- or magnetoencephalography identify nodes with sensors. This introduces spatial scales possibly unrelated to the actual system organization, and potentially affecting the network's topological properties [82,83]. Furthermore, it is not clear whether the functional networks based on surface recordings actually reflect the topology of the underlying network of neuronal sources [84,85].

Multiple electrode recordings tend to overestimate the true network small-worldness of the underlying network, as each sensor picks up many sources at small scales. The small number of sensors constrains the sampling on large scales [86], whereas the imperfect sampling may impede the detection of scale-freeness [81].

In functional magnetic imaging studies, the main issue associated with node identification is that of delineating functionally separated brain units. This task, which goes under the name of *parcellation*, exposes a series of very general issues related to the representation of a functional space, and to the correct definition of the corresponding tools provided by complex network theory to account for the internal organization of such spaces.

Parcellation may identify nodes with either anatomical landmarks, or locations in the brain volume, or else with peaks in functional activation [87]. While the choice of the parcellation technique may not affect whether or not certain global topological properties such as small-worldness are actually detected, it may nonetheless have an influence on their quantitative estimates [88].

However, on the one hand, methods based on anatomical landmarks rest on the controversial isomorphism between anatomical and functional spaces [89]. On, the other hand, using functional landmarks may lead to a fluctuating number of nodes, and care should be taken when comparing the associated topologies.

In general, parcellation of neuroimaging data typically yields very high-dimensional data sets. Principled dimension reduction should preserve physiologically relevant information and functional organisation rules. However, the anatomo-functional space is often segmented with partitional clustering methods [90]. This class of methods typically involves two unrealistic assumptions: the same region cannot simultaneously participate in different functional units. At the same time, all parts are typically forced into belonging to at least one cluster. In addition, most methods used to detect modularity are not robust with respect to the presence of well separated scales [91], and are therefore ill-suited to reconstruct organizational principles at different levels.

### (ii) Defining links

Functional links usually reflect statistical relationships between activity recorded at different brain sites or sensors.

How different connectivity metrics affect the topological properties of the resulting networks and how to elect the most appropriate metric of brain activity out of the great number of available ones are still poorly understood issues. Because even slight changes in connectivity patterns may result in large changes in the measurements made at a particular node, these factors are likely to have a non negligible impact even at macroscopic scales.

Another important issue is how to transform an all-to-all connected clique into a functional network. However carried out, this generally involves setting a threshold value either *a priori* [21], or after examining a range of values [92,93], or through an adaptive process [19], e.g. by choosing its maximal value keeping the network connected [62]. A qualitatively different strategy consists in selecting the threshold level that optimizes some criterion, e.g. data classification [94].

Setting a threshold has several interrelated, potential consequences. First, too high a threshold can prevent the convergence of the sample distribution to the true asymptotic one and therefore the emergence of the corresponding macroscopic property. The percentage of considered links may also not be the one which optimizes data classification based on network properties, and where classification is robust to fluctuations in network parameters [94].

Furthermore, thresholds are filters biasing the analysis towards given scales and corresponding topological properties, damping the effect of other ones. The multiscale nature of brain activity suggests that no filter is optimal and that choosing a threshold value only determines what properties the analysis is going to shed light upon, as each network metric is strongly associated with a preferred link density. For instance, triangular motifs cannot appear in very sparse networks, while unconnected triangles disappear in very dense networks. Similarly, hub-based structures fade out for very high link densities. At macroscopic scales, brain activity may appear hierarchically organized into modules with *large-world* self-similar properties, while the addition of only a few weak links is enough to turn the network into a non-fractal and *small-world* one [27].

### (c) Interpreting network properties

Once a network is constructed, one needs to interpret the meaning and significance of the properties one wants to characterize it with.

Connectivity measures should not be taken to automatically reflect the presence of specialized structure, due to the strong influence exerted by geometry on connectivity matrices [95]. Most topological properties typically attributed to brain structure, including modularity and hierarchy, can be seen in strictly uniform, locally connected two-dimensional spaces. Given the prominent role played by geometric constraints in the brain, this is a potentially serious problem, which implies that the role of geometry must be discounted before interpreting observed topologies by analogy with known results from different fields in which network theory has been used.

There is no clear relationship between connectivity and transfer or processing of information. The relationship between information processing capacities and topological network properties has been investigated theoretically [53]. However, the existing literature typically quantifies the information contained *in* the observed network, while the actual information processed or transferred by the underlying system is often difficult to quantify due to technical limitations.

The definition of some constructs of network theory (e.g. efficiency, robustness or vulnerability) was dictated by the specific constraints of the physical contexts they were designed to describe (e.g. internet, WWW, social networks), but may not be appropriate in the case of brain activity.

For example, in the framework of complex network theory, the term *efficiency* is defined as the inverse of the number of steps needed to reach one node from any other one in the network [52]. Thus defined, *network efficiency* qualitatively differs from the usual definition of efficiency, which relates to the way the system takes advantage of its resources to perform a given task, and should therefore not be equated to it.

Other examples of these risks are the evaluation of robustness and vulnerability. Both parameters are traditionally obtained by targeted or random removals of the network nodes and/or links [57]. The effects of node and link deletion in the network parameters are reasonable in networks like the Internet, where all routers and servers have similar functions [96]. Nevertheless, it is adventurous to extrapolate these techniques to brain networks, due to the fact that each brain node is different, and that the deletion of a node or link may have qualitatively different consequences from those predicted by the topological robustness. For example, the removal of a crucial but poorly connected node may lead the whole brain network to fail when performing a cognitive task.

On the other hand, some network concepts have been borrowed from other domains of application, disregarding the conditions under which they are valid. Synchronizability of a functional network is a paradigmatic example. This network parameter has been used to evaluate whether a complex network is able to synchronize or not [97], and it has also been translated to functional brain networks [19,62]. Synchronizability relies on the spectral properties of the Laplacian matrix associated with the functional network. Nevertheless, this parameter requires all nodes of the network to be identical systems, something that is far from being the case of the brain. Furthermore, it refers to both phase and amplitude synchronization of the full system (i.e., complete synchronization) [63,98], a kind of synchronized behaviour never reported in biological systems.

### (d) Considering the true dimension(s) of networks

Functional networks are continuously evolving, even at rest. To capture the behaviour of functional networks, time must be included in the analysis. Most existing studies describe functional networks in terms of steady-state (topological or dynamical) network properties averaged over a given experimental condition. Averaged steady-state networks will inevitably tend to approximate anatomical ones, as anatomical networks are functional ones averaged over an infinite time window.

However, functional networks are inherently transient, as the time in which functional links reconfigure is typically orders of magnitude faster than the length of neural processes. When the duration of a given phenomenon is many orders of magnitude larger than that of changes in the wiring, the temporal dimension and its structure cannot be neglected, whether the process is stationary or not.

Very few studies have dealt with how these networks emerge, evolve and disappear [19,99-101]. Due to the presence of delays, functional networks are not only spatially extended but also temporally non-local. However, connectivity is typically evaluated locally in time, and with a single characteristic scale.

Moreover, while network theory is used in recognition of the existence of non random *spatial* structure at a variety of scales, brain activity also has a multiscale *temporal* structure. Importantly, the generic presence of complex fluctuation properties such as scale-freeness and long-range temporal correlations, and rich non-trivial hierarchical [102 and references therein] and ordinal [103] temporal structure indicates that activities at various scales are not separable, so that describing brain activity boils down to accounting for the rules governing their relationships. The relationship between different time scales is typically forgotten or explicitly avoided as there is as yet no standard methodology to quantify the connections between temporal scales [104]. Failure to account for the non-random structure associated with the complex generic properties of the temporal scales of brain fluctuations leads to missing or distorting temporally non-local structure, and does not help understanding the complex interactions among structures unfolding at very different characteristic time scales [102].

An important and related issue is to determine what quantities can be averaged together and how. From a statistical mechanics viewpoint, the generic presence of properties such as modularity and small-wordness makes the brain a disordered system. The disorder found in the brain can be thought of as an externally given background and should therefore be considered as *quenched disorder*, i.e. the parameters defining its behaviour are random variables which do not evolve with time, with modules playing the role of impurities. While there is some indication that quantities measured for such system may indeed be *self-averaging* [105], indicating that statistics are improved by increasing the sample, it is not immediately evident that the distribution of impurities does indeed obey the equilibrium distribution.

### (e) Evaluating results

#### (i) Discriminating important features

Once a series of network metrics are calculated, neuroscientists face the arduous problem of understanding what properties are important [94].

Statistical inference often relies on at best a few metrics such as the path length, or clustering coefficient [56,106,107]. Significant differences in graph metrics can highlight differences between groups, but incur problems related to multiple comparisons. More importantly, lead to dramatical information losses as a result of the reduction of a complex system to a set of scalars [9]. In addition, the standard statistical analysis does not provide a principled way to favour one property over another, neither does it account for the relationship between different metrics, which remains unexplored.

#### (ii) Reproducibility, sensitivity and specificity

The extent to which measured properties actually describe the system and are specific to it remains unclear.

In spite of the great number of studies reporting topological differences between the functional networks of patients suffering from a variety of pathologies, the sensitivity and specificity of such metrics may not be sufficient to be clinically useful or have an effective diagnostic value [9].

The mapping from *microstates*, represented by observed functional network structure, to *macrostates*, represented by the corresponding ability to perform a given task or by a given pathology, may be extremely unpredictable. Rather dramatic changes in the former may turn out to be neutral, failing to translate into appreciable functional change, which instead may occur in association with seemingly small ones. This complexity is to be expected from a network where each node represents a degenerate or conversely a pluripotential system, respectively characterized by a many-to-one or one-to-many structure–function relationship.

Finally, inconsistent results have occasionally been highlighted [9]. For instance, epilepsy has been associated to both decreased [108] and increased [109] path length with respect to normal control groups. However, studies analysing the reproducibility of network parameters are scarce and no clear picture emerges in this respect [110-112], partly as a result of the lack of understanding of intrinsic brain response *consistency* [113], and adaptation, and their role in shaping network topology.

## 4. The great leap forward

How can complex network theory move up gears, and start delivering the goods that the neuroscience community expected of it?

In the remainder, we propose some ways and conditions through which this can be accomplished. These include calling upon some already existing conceptual and technical aspects of complex network theory that have not yet been resorted to by neuroscientists, and proposing tailor-made metrics consistent with known properties of functional brain properties and, as a result, of a wide class of complex adaptive systems.

Whether all, some, or even none of the proposed recipes bears fruit or not, the most important goal of this section, as of the paper as a whole, remains that of promoting a constructive debate on the future of complex network theory in neuroscience.

### (a) Taking full advantage of graph theory

#### (i) Multiscaling in space, time, and phase space

Functional brain networks have an inherent spatio-temporal dimension. A time-varying description of functional networks naturally leads to a *multilayer* network representation [114], with layers labelled by time.

Time-varying and multilayer networks involve a basic reformulation of most of complex networks' founding concepts, from topological properties as basic as distances, to community structure and modularity, small-worldness, etc. [114-116]. This specific field of complex networks analysis is still in its infancy and could benefit from the experimental results coming from functional brain networks.

The most appropriate mapping should take into account both the spatial and temporal scales, equipped with their respective structure. Two complementary approaches constitute dual cuts into this space: on the one hand, considering connectivity at different time scales helps unveiling hierarchical neural communities [117]. Likewise, the ordinal and hierarchical temporal structure can be explored by sweeping the spatial scales.

The brain can be considered as a complex many-body system, many aspects of which evolve with the resolution scale at which it is observed [118]. In other words, to capture this essential principle of brain functioning methods are needed that are able to deal not only with activity at one or many particular scales, but also with the relationship across scales. Out of the many possible solutions to this fundamental problem, the *renormalisation group* appears as a general paradigmatic method providing a compact representation of the relationships across scales [119].

A renormalisation group is in essence a dynamical system, where time axis is represented by the logarithm of the scale factor, describing the evolution of models of a system in a model space, as the space of models is mapped into itself, through coarse-graining to longer lengths. The evolution of scale-dependent parameters under coarse-graining can generally be expressed in terms of differential equations for the probability distribution function. In the case of networks, this can be accomplished by covering the network with boxes of a given size and then replacing each box with renormalized *supernodes* [120].

Insofar as it is a dynamical system, the renormalization flow can be characterized by its fixed points and their stability. The fixed points express the properties that are conserved as scales are varied and details at small scales are lost. The various asymptotic behaviours of the system emerge as scale-dependent collective phenomena. What particular behaviour, out of the many possible ones, is attained by the system under the action of coarse graining depends on the initial parameter values' location within the basins of attraction of the fixed points. Power-law and hierarchical structure are two of the classes of asymptotic behaviours that can emerge as an out-of-equilibrium system is coarse-grained [118].

The renormalization flow helps representing the various observable network configurations as the phase space of a dynamical system, i.e. the abstract space of all possible states brain activity can take, bridging the gap between functional networks at scales as far apart as those of perceptual phenomena, of brain plasticity or aging, and even of evolution. The renormalization group approach can be seen as a natural method to tackle the problems of describing, modelling, and in some sense even predicting multiscaleness in the brain. Renormalization theory helps relating models of the same system at different scales or grouping models of different systems exhibiting the same large scale behaviour.

While the standard renormalization procedure looks at the evolution of effective parameters and, as it were, at the information conserved by the flow, quantifying the information lost as the look it progressively zoomed out of the system helps characterizing mesoscale properties, which tend to vanish for diverging time and network size, but are observable at the spatial and time scales typical of functional brain activity [53,121].

(ii) Topology – dynamics

While nodes are generally taken to be static objects, it is possible to endow them with some evolution rule [5]. Given the typically oscillatory nature of brain activity, networks reconstructed from brain activity boil down to a set of oscillators (weakly) coupled according to a certain topology.

Adding dynamics allows resorting to the rich repertoire of tools of nonlinear time series analysis [122]. For instance, one can derive properties of the dynamics such as equilibria and their stability, as well as other fundamental dynamical and geometrical properties of the phase space associated with the dynamics, and the bifurcations it may undergo as some control parameter is being varied. This naturally leads to the definition of a *dynamical robustness* and *vulnerability*. Contrary to *topological* robustness, where typically one assesses the evolution of the largest connected component, as nodes or links are deleted using some strategy, when considering dynamical robustness, the critical variable is a dynamical network property, such as synchronization. An important difference is that both perturbing fields and their consequences can be continuous and smooth are therefore endowed with more general and better defined properties than the all-or-none lesioning considered in topological robustness studies.

Dynamics can also be introduced in a slightly different way. Complex networks have an irregular wiring that naturally lends itself to a statistical description. The equilibrium statistics of networks can be described by a partition function defined as a sum over all graphs with a fixed number of vertices and links, from which the potentials describing the system's thermodynamics [123]. Network properties can be used as order parameters, the behaviour of which can be monitored as the value of a variable controlling the system, e.g. some network variable or a cognitive task, is being manipulated. Critical phenomena such as structural phase transitions or the emergence of scale-free architectures can then be assessed [124].

Furthermore, one can study the interplay between the dynamics of nodes of the network and that of the network topology, which can itself be regarded as a dynamical system [125]. For instance, the temporal structure, e.g. burstiness and intermittency, influences the spreading of information in a network [126], and the relative time scale of topology, intermittency and of its exponential tail influence the relaxation time of the underlying process to its stationary distribution [127]. How observed dynamical properties of nodes (which can take any spatial and temporal scale) relate to the topological network properties at all scales, and how both translate into observed function (e.g. the proficiency level in the performance of a given task) constitute research avenues in their own right, which demand to be explored.

(iii) Beyond isolated networks: interacting and competing networks

Networks do not live in isolation. Instead they generally interact with other networks. It is then interesting to study previously separated networks that become interdependent as links uniting them are formed. While to some extent surrendering their independence as a result of interaction, each of these networks retains its own identity.

*Networks-of-networks* [128] present a very rich and surprising phenomenology, often running counter the intuitions afforded by results obtained for isolated networks, for robustness [129], centrality [130] or synchronization [131]. For example, the evaluation of the importance of a node in a network has traditionally been quantified by means of the eigenvector centrality [76,132-134], a measure based on the spectral properties of the functional network. However, it has recently been shown that the existence of interacting sub-networks (or modules) and the way they interact strongly determine the distribution of centrality within the whole network [130]. This ultimately means that the reorganization within a sub-network affects the importance of nodes belonging to other sub-networks. Similarly, recent results show that the way network modules are interconnected also determines the ability of the whole network to synchronize [135], a fact that influences the analysis of functional networks.

Clearly, this reformulation has to be translated to brain functional networks, where it is crucial to understand how functional sub-networks subserving different cognitive functions interact and compete between each other, how their efficiency is altered or diminished as a result of interaction, and how processes such as synchronization are favoured under certain connectivity patterns.

### (b) Generalizing the use of network representations

Functional brain activity is typically represented in a space isomorphic to the anatomical one, with nodes reflecting anatomically-related units, and links a connectivity metrics. However, network theory could be used to describe functional brain activity in rather different ways. This may in part be motivated by the fact that connectivity may not be the best descriptor of functional activity. Function may for instance emerge from a collective property independent of connectivity [89].

Network representations of functional brain activity need not be isomorphic to the topology of the brain. Network theory may be used to describe the *phase space*. One way to achieve this is to conceive of brain activity as a random walk in a high-dimensional space, and to use network theory to model the way the space is being visited by the dynamics. Brain dynamics has been shown to be weakly non-ergodic [136], a condition where the whole phase space is still accessible, but the time to visit certain regions may be much longer than typical experimental ones [137]. Because complex networks are strongly disordered systems, where fluctuations of structural characteristics may far exceed their mean values [124], the inhomogeneity of functional brain activity's phase space could be endowed with a network representation, with microscopic dynamics restricted to nodes and links [138]. The effects of cognitive tasks or brain damage may then be gauged in terms of changes in macroscopic topological and dynamical properties of the functional space.

More generally, the space of functional brain activity may take arbitrarily complex forms, comprising information with heterogeneous dimensionalities and possibly incommensurable natures. Imagine for instance that available data would document different aspects of activity of a given subject. These data may come in the form of a time series (e.g. an EKG recording), but also of static scalar vectors (e.g. blood tests; or behavioural neuropsychological scores), or a matrix mapping different values in space (e.g. a CT scan imaging). While these tests account for a unitary underlying system, from a data analysis view-point, understanding this information set as a system may represent an challenging step. Overcoming this "perceptual" stumbling block would allow generalizing graph theory to a class of contexts that are usually not thought of as systems. In [139] it was shown how such systems can be represented as networks, called *parenclitic networks*, where nodes represent features, and links quantify deviations between two features and their typical relationship within a population. The information on the structure of this generalized functional space is ultimately embedded in the topology of the reconstructed network.

Finally, networks may be thought of as a rich convenient space onto which time series and other data formats can be transformed, the mapping being bijective under rather general conditions [140]. Thus, network analysis can be used to distinguish different dynamic regimes in time series. Conversely, time series analysis can map the system's network statistics into dynamical properties.

### (c) A neuroscience-inspired graph theory

While showing a certain degree of universality and independence [33], each system may possess idiosyncratic properties. On the other hand, complex network theory is a branch of applied physics: its tools and the quantities it measures are bound to somehow reflect some of the specific characteristics of the system it is meant to describe. Historically, complex network theory was developed to model systems in many ways qualitatively different from biological systems in general and from the brain in particular.

Some fundamental elements of neural function, viz. inhibitory connectivity and feedback loops, have not yet been incorporated in the standard toolkit of functional network description. While generally difficult to capture with standard non-invasive neuroimaging techniques, and not mapped in straightforward way by negative links [9], inhibition should nonetheless be incorporated into network models of functional brain activity. A similar remark befits feedback loops.

For complex network theory and neuroscience to meet each other's needs a few other adjustments seem desirable. For one thing, it would be useful to integrate the fact that the brain possesses qualitatively different nodes, be they neurons or entire brain regions. For other basic concepts, e.g. that of distance, neuroscience should promote alternative definitions, at least when considering a functional space isomorphic to the anatomical one (as opposed to a phase space representation). Likewise, community structure should be redefined in such a way as to account for the possibility for a given neural assembly to pertain to different communities, possibly at different spatial and temporal scales.

Network properties should reflect the fact that the brain is a complex adaptive system. This requires a clear understanding of how functional networks respond to external stimuli at various spatial and temporal scales, or damaged brain networks adapt, after both permanent neurological damage and, at faster time scales, e.g. following epileptic seizures, and become active again [141]. Parameters measuring brain *adaptiveness*, including topology-dynamics interactions should be proposed. The *evolvability* of a network, i.e. the continued propensity to adaptive innovation [142], may be estimated by quantifying *navigability* within the network representation of the system's phase space, i.e. the system's ability to find any given region of its phase space starting from any other one [143].

In addition, robustness should be defined in a functional, rather than structural way, accounting for the complex relationships between robustness, complexity, and evolvability. Nested time-scale dependent notions of robustness, defined for different levels of organisation, which allow reconciling the conflicting requirements for robustness and adaptability should be given a network translation [144].

Finally, research should strive to bridge the gap between information encoded *in* the network, i.e. the information contained in the structure that is analysed, and that encoded *by* the network, i.e. the information actually treated or transferred by the brain [53]. The first step may consist in acknowledging that communication in brain networks can take place through many more routes than the shortest paths. To this end, several notions of *communicability* have been introduced [145]. These measures take into account all possible routes between two nodes, assigning smaller weights to longer ones. More fundamentally, a representation is needed of the way the system stores and processes information. This requires going beyond the classical statistical mechanics approach, which derives macroscopic consequences of microscopic dynamics, but does not provide information on how the system stores and processes information, and adopting a *computational mechanics* one, producing causal models capable of generating the statistics of observed time series and therefore the underlying stochastic process [146].

### (d) Broadening objectives

While complex network theory provides an impressively rich set of tools to characterize brain functional activity, neuroscientists' objectives go beyond the pure description level and would benefit from tools that are able to address some of their fundamental demands: classifying patients or experimental conditions, understanding the aetiology of observed connectivity patterns, modelling activity in as complete a way as possible, to eventually not only be able to forecast and control it, but also to steer it to desirable states.

#### (i) From comparison to classification and categorization

If network properties genuinely describe functional brain activity and its modulations under given conditions, e.g. cognitive tasks or neurological pathologies, it should be possible to use them to discriminate various activity regimes associated with these conditions.

One principled way to overcome this limitation involves assessing what network properties optimize a given task, for instance classifying experimental samples corresponding to different experimental conditions. The amount of information codified in each network can be approximated by the success score achieved in a classification task, where a model is trained to identify subjects belonging to the two considered classes [94]. Not only does this strategy allow identifying the combinations of properties obtaining higher classification scores, but it also affords a quantitative assessment of the degree to which these properties actually discriminate between different experimental conditions. This strategy is by no means confined to classification tasks; for instance, it could conceivably be applied to modelling and predicting certain types of behaviour.

Another strategy may consist in trying to define networks along some metrics that would allow calculating distances between them. For example, distances between pairs of networks, and ultimately network taxonomies, can be developed using the response function of community structure to changes with scale [147]. On the other hand, the definition of appropriate *morphospaces*, i.e. phenotype spaces with defining quantitative traits as axes [e.g. 148] should help refining the properties that are selectively modified by the experimental conditions under study and would allow comparing along common traits configurations associated with qualitatively different functional properties.

Statistical mechanics techniques can be applied to brain networks to quantify the statistical significance of empirically observed properties [95]. For instance, an observed network can be thought of as a specific instance either of a particular network evolution, or of an ensemble of networks, subject to some (e.g. functional) constraint [149]. Networks can be characterized by considering a series of *randomized network models*, i.e. null models of real networks conserving some of their, e.g. degree distribution, or community structure. Network ensembles with a given sequence of values of a given property fall into the class of hidden variable models, where the hidden variable is represented by the elasticity of the topology to changes in the properties that are being optimized. The role of each structural feature in a given network can be measured by the network ensemble's entropy, i.e. the normalized logarithm of the total number of networks belonging to the ensemble. This may allow building pseudo-metrics and as a consequence measuring distances between different experimental conditions.

Network topologies [33] and, to some extent, network dynamics [150] present universal properties. Observed properties can be assigned to *universality classes* using the renormalization group theory [120]. Universality classes are the basins of attraction of fixed points of renormalization flows, the points within which have the same properties on large scales. Furthermore, the surface comprising the models flowing into the same fixed point separates the space into different phases. Because the functional space is not always easy to navigate, universality classes and renormalization flows are important tools for partitioning the phase space, thus lending an important hand in comparing and classifying observed networks.

#### (ii) Modelling and forecasting

Arguably the first step into understanding the mechanistic properties of a given observed phenomenon is defining its aetiology. Although the notion that the topology of biological networks can provide insights into its functioning principles is debated [151,152], different types of networks are likely to be generated by different mechanisms, and their topology may give clues as to the mechanisms that created them. Network topology may contain information on the design principles of biological networks and therefore provide some clues into the dynamical evolutionary processes that generated these networks [153].

One may want to understand the selection forces shaping functional activity at evolutionary time scales, or the rules generating a given observed steady-state or a time-varying functional pattern at far shorter time scales [146]. The fundamental forces that shaped human brain network topology at evolutionary time scales remain poorly understood, and only few computational studies explored the role of factors including energetic costs, communication efficiency, and dynamic complexity [4]. A similar dearth plagues our current knowledge of functional activity at faster time scales [42,43,154,155].

Characterizing stylized facts, i.e. structural characteristics that would hold for a diverse collection of instruments, experimental conditions, and time scales may ultimately enable modelling observed time series for a range of scales, and for instance predicting the next steps for a given sequence of data. This in turn may supplement existing comparative statics approaches to the appraisal of the functional potential of brain systems for future learning [22], prior to or following brain damage.

#### (iii) Controlling and targeting of functional brain properties

So far, we have seen that the way the brain responds to an external field can be endowed with a network representation, for instance in terms of structural and dynamical vulnerability. Various recently proposed methods may help taking a step further, i.e. understanding how to perturb the system in desirable ways, typically by acting on a limited number of nodes.

It is for instance possible to *control* a functional network [156], i.e. to stabilize the system within a dynamic regime it would not naturally reach, or to *target* a desired dynamical state [157], i.e. to steer the system towards a goal dynamics which would naturally be achieved starting from a different initial condition. The former may for example be used to keep network dynamics away from a pathological range and to stabilize it within a healthy one.

This may represent a qualitative advance in the treatment of various pathologies for which therapeutic strategies do not fully take into account the network structure used to represent them. One notable example is represented by the standard surgical treatment of pharmacologically intractable epilepsy. The standard surgical approach still consists in resecting or disconnecting epileptic foci. The fact that a significant minority of patients continue to experience seizures after surgery, particularly in the presence of multiple epileptic foci, suggests the inadequacy of this surgical strategy. While a network characterization of epileptogenesis has recently emerged [158], a surgical strategy based on such an understanding may help overcoming the current shortcomings.

Importantly, the aetiology of a given pathology need not be network-like for network control to possibly be effective. For instance, while Parkinson's disease's causal factors originate in a well-identified and circumscribed brain region, its consequences affect the functioning of various circuits, and its surgical control via implanted stimulators could target a global network dynamics rather than a unique well-localized brain region.

Targeting techniques could also find interesting applications in cognitive neuroscience. For example, as network-based descriptions of various learning processes get more accurate [22], it may become possible to shorten the learning path by targeting desired network dynamics. Transcranial magnetic stimulation, biofeedback, or pharmacological manipulation could represent experimentally viable non-invasive ways to drive brain dynamics or, at least, to study the ability/resistance of the brain to be driven.

Finally, considering the adaptive self-organizing nature of brain activity, one riveting research avenue would imply engineering adaptive rules such that a given topology self-organizes into a desired state, with desirable dynamical and functional properties.

## 5. Conclusions

Will complex network theory ever bring about a revolution in the field of neuroscience?

We have tried to argue that there are strong reasons for that to occur, for not only has complex network theory got the potential for vastly increasing the ability to describe the brain as a complex biophysical system and to understand its basic organization principles, with respect to previous methods, but it may also provide appropriate tools for its targeted manipulation, with obvious applications in the clinical and cognitive domains.

Exploiting complex network theory's full potential will suppose a few conceptual quantum leaps. The statistical mechanics assumptions representing the backbone of complex network theory and their conceptual and methodological implications will have to be interiorized. At the same time, some of its intrinsic limits will need to be acknowledged and overcome. Neuroscience will have to both resort to hitherto unexploited existing network tools, particularly accounting for dynamical aspects of brain activity, and to stimulate fresh theoretical effort, so as to produce network constructs better catering for its specific needs, instead of importing wholesale and readymade concepts originally meant to describe systems in many ways qualitatively different from the brain.